\documentclass{article}
\usepackage{spconf,amsmath,graphicx}
\usepackage{cite, array, booktabs}
\usepackage[hidelinks]{hyperref}
\usepackage{amssymb}
\usepackage{multirow}
\usepackage{xcolor}
\usepackage{tablefootnote}
\usepackage{adjustbox}

\title{SpeechLMScore: Evaluating speech generation using\\ speech language model}
\name{Soumi Maiti$^1$, Yifan Peng$^1$, Takaaki Saeki$^{1,2}$, Shinji Watanabe$^1$}
\address{$^1$Carnegie Mellon University, USA, $^2$The University of Tokyo, Japan}
\begin{document}

\ninept
\maketitle
\begin{abstract}
While human evaluation is the most reliable metric for evaluating speech generation systems, it is generally costly and time-consuming. Previous studies on automatic speech quality assessment address the problem by predicting human evaluation scores with machine learning models.
%and achieved a high correlation with human evaluation. 
However, they rely on supervised learning and thus suffer from high annotation costs and domain-shift problems. We propose \textit{SpeechLMScore}, an unsupervised metric to evaluate generated speech using a speech language model.
%which is analogous to language-model-based scoring in text generation.
SpeechLMScore computes the average log-probability of a speech signal by mapping it into discrete tokens and measures the average probability of generating the sequence of tokens. Therefore, it does not require human annotation and is a highly scalable framework. Evaluation results demonstrate that the proposed metric shows a promising correlation with human evaluation scores on different speech generation tasks including voice conversion, text-to-speech, and speech enhancement.

%1) the proposed metric shows a promising correlation with human evaluation scores on the VoiceMOS challenge 2022 dataset, and 2) the scores with SpeechLMScore are also correlated with the noisiness in speech. 

\end{abstract}

\section{Introduction}
\label{sec:intro}
Human evaluation is the gold standard for evaluating different speech generation systems including speech synthesis, voice conversion, and speech enhancement. It is typically carried out with subjective listening tests such as Mean-Opinion-Score (MOS) tests on 1-5 Likert scale, where a participant listens to an audio sample and provides judgement. Such subjective tests are time-consuming, costly and hence are not scalable. % to large numbers of systems or files. 
%In addition, MOS are relative scores not absolute measures, i.e they are scores of a sample compared to the other samples present in the test~\cite{lemaguer22_interspeech}.
% In addition, MOS is a relative score to the test, i.e. humans generally rate the speech relative to the other samples they listen to.  
In contrast, objective evaluation metrics are easy to use for large numbers of systems and can provide faster results but these objective metrics are generally not highly correlated with the human judgement score. For example text-to-speech (TTS) objective metrics like Mean-Cepstral-Distortion (MCD)~\cite{fukada1992adaptive} or F0-root mean square error (F0-RMSE) poorly correlate with human evaluation as they only measure certain acoustic similarity with a reference speech. Similar phenomena are observed for speech enhancement objective scores like PESQ~\cite{rix2001pesq}. 
%Therefore, even though listeners always do not agree with the exact score for the same speech, listening tests remains the preferred method of evaluating different speech systems.

To tackle this problem, recent studies~\cite{lo2019mosnet, sellam2022squid, reddy2021dnsmos, cooper2022generalization} treat the speech evaluation as a regression or classification task and train a neural network to predict MOS. Such models require large-scale listening tests with multiple participants for supervision. For example, recent work~\cite{sellam2022squid} conducts listening test with 1.3M samples for multilingual MOS, \cite{huang2022voicemos} collected human evaluation on 7K samples. 
%and train a neural network to predict the MOS. Such supervised models are trained on datasets collected with very large-scale listening tests in one or more domains with multiple participants and use averaged MOS as a ground-truth label. 
In addition to the high cost of collecting human judgements, such supervised models also risks poor generalization to new tasks and domains. In this paper, we propose to evaluate speech generation by measuring the likelihood with a speech language model, instead of predicting non-absolute human scores. In recent years, speech language models~\cite{lakhotia2021generative, kharitonov2021text, borsos2022audiolm} are introduced for the speech continuation task or unconditional natural speech generation. 
%Such a speech language model uses encoder to map speech into discrete speech units and then train an autoregressive language model (LM).
Here, we argue that we can also use such speech language models for measuring likelihood of another speech sample.

More specifically, this paper proposes an unsupervised speech evaluation metric, \textit{SpeechLMScore} based on speech language models. SpeechLMScore measures the likelihood of a speech sample to natural speech using a pretrained speech-unit language model.  
SpeechLMScore is non-intrusive i.e., it does not require the reference speech for measuring similarities unlike some objective measures MCD, F0-RMSE or PESQ. More importantly, SpeechLMScore is an unsupervised metric, hence it does not need explicit human MOS labels. SpeechLMScore is not optimized for a particular task and hence is more generalizable than other metrics.
%We demonstrate that using pretrained unit-speech LMs we can build such metric easily without any extra training costs or adaptation. Further, 
We show that SpeechLMScore is highly correlated with different speech evaluation tasks like voice conversion, TTS, and speech enhancement. Especially, SpeechLMScore shows higher correlation in mismatched domains compared to supervised MOS prediction models. Furthermore, we show that using duration information of speech tokens in speech language model, correlation is improved. 
We will open-source SpeechLMScore evaluation metric, speech language model training code and pretrained speech language model using ESPnet toolkit\footnote{\url{https://github.com/ESPnet/ESPnet}}.

\begin{figure}[!t]\label{fig:ulm}
\centering
\includegraphics[width=0.45\textwidth, trim={1.42cm 1.0cm 3.0cm 0},clip]{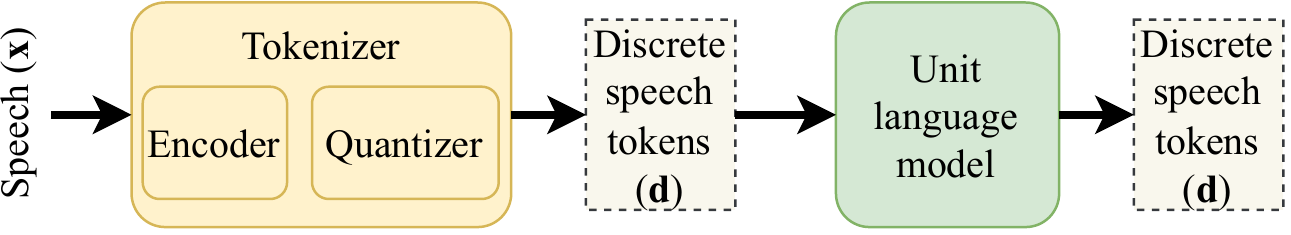}
\caption{Speech Language Model~\cite{lakhotia2021generative} using discrete speech units.}
\end{figure}

\section{Related Work}
\label{sec:rel_work}
% text generation evaluation with language modelling- BARTScore, BERTscore, MoverScore

In the field of natural language generation, unsupervised automatic evaluation methods with language models~\cite{zhao2019moverscore,zhang2019bertscore,yuan2021bartscore} have been proposed.
Unlike traditional metrics such as BLEU~\cite{papineni2002bleu}, these evaluation methods can capture more semantic information inherent in the language model, leading to higher robustness and correlation with human evaluations.
In recent years, some studies perform audio generation as neural machine translation~\cite{hayashi2020discretalk} or language modeling tasks~\cite{lakhotia2021generative,borsos2022audiolm} by treating acoustic signals as discrete tokens, opening up a new frontier in audio generation.
Therefore, our study naturally extends the language-model-based scoring to a speech generation task, which is a promising and highly scalable framework.

Automatic speech quality assessment methods for speech generation tasks such as speech synthesis~\cite{patton2016automos,lo2019mosnet,huang2022voicemos} and speech enhancement~\cite{reddy2020dns} are actively being studied.
Recently, transfer learning from self/semi-supervised representation learning models~\cite{cooper2022generalization, saeki22c_interspeech,sellam2022squid} has shown remarkable prediction accuracy and generalization ability.
However, these models are trained with supervised learning and thus require large-scale listening tests to collect a large number of target labels.
Therefore, previous work aims to learn MOS prediction models with non-matching references for the application to out-of-domain data~\cite{manocha22c_interspeech}.
In contrast to the above studies, our approach performs scoring with a speech language model and allows automatic evaluations without any human annotations.

\begin{figure}[!t]\label{fig:speech_lm_score}

\begin{minipage}[b]{1.0\linewidth}
  \centering
  \centerline{\includegraphics[width=0.8\textwidth]{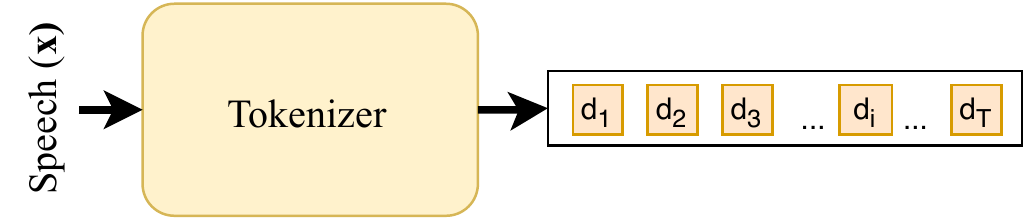}}
%  \vspace{2.0cm}
  \centerline{(a) }\medskip
\end{minipage}
\begin{minipage}[b]{1.0\linewidth}
  \centering
  \centerline{\includegraphics[width=0.95\textwidth]{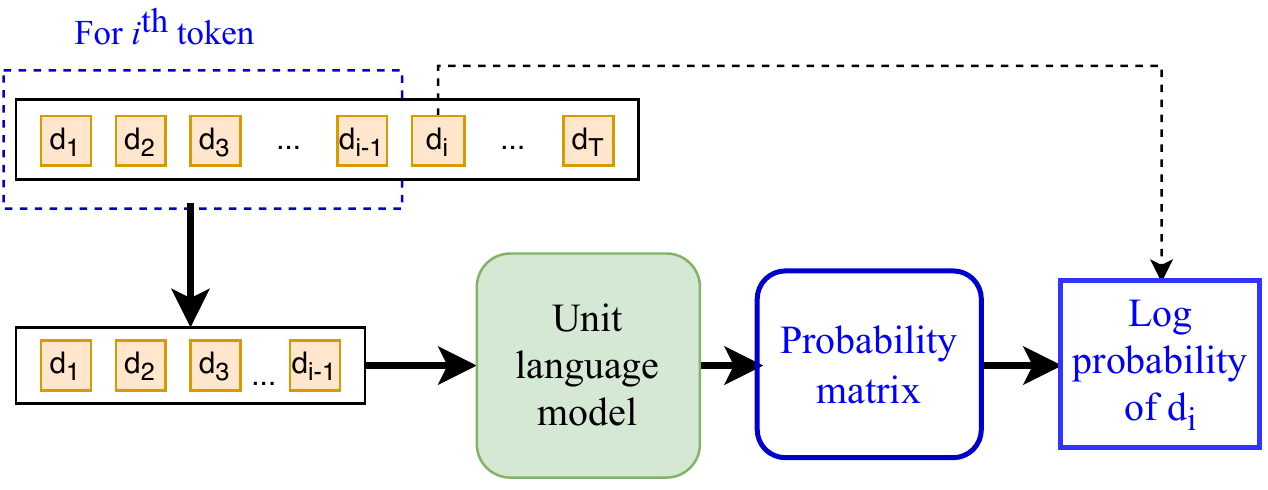}}
%  \vspace{1.5cm}
  \centerline{(b) }\medskip
\end{minipage}
\caption{Computation of SpeechLMScore: (a) Given speech waveform $\mathbf{x}$, encoder generates discrete tokens (b) For each token $d_{i}$, a unit language model generates probability matrix over vocabulary $V$ given all previous tokens. % and log-likelihood of current token is calculated. 
Finally, SpeechLMScore($\mathbf{x}$) is calculated as the average log-likelihood of all tokens.}
\end{figure}

%%%%%%%%%%%%%%%%%%%%%%%%%%%%%%%%%%%%%%%%%%%%%%%%%%%%%%%%%%%%%%%%%%%%%%%%%%%%%%%%%%%%%%%%%%%%%%%%%%%%%%%%%%%%%%%%%%%%%%%%%%%
\section{Method}
\label{sec:method}
%A speech language model generates the probability distribution over a set of discrete speech tokens. We propose that we can evaluate the quality of a speech signal from a speech language model by computing average log-likelihood.

Given a $t$-length speech signal $\mathbf{x}=\{x_1 \cdots x_t\}$, $x_t \in \mathbb{R}$, we can compute a joint probability of speech signal using chain rule as
\begin{equation}
    p(\mathbf{x})= \Pi_{i=1}^{t} p(x_i| x_{<i}),
\end{equation}
where $p(x_i| x_{<i})=p(x_i| x_1\cdots x_{i-1})$.
% We can also compute log-likelihood as,
% \begin{equation}
%     \text{log-likelihood}(\mathbf{x})=\sum_{i=1}^{t} \log p(x_i| x_{<i})
% \end{equation}
A speech language model $\theta$, generates the probability distribution of speech $\mathbf{x}$ as
\begin{equation}
    p(\mathbf{x} | \theta)=\Pi_{i=1}^{t} p(x_i| x_{<i}, \theta).
\end{equation}
We propose to estimate quality of speech $\mathbf{x}$ with average log-probability of speech computed using a speech language model.

%SpeechLMScore estimates quality of speech $\mathbf{x}$ with a probability from the speech language model.
%joint speech likelihood ($p(\mathbf{x})$) given a speech language model $\theta$.

\subsection{Speech language model}\label{ssec:speech_lm}
A speech language model was proposed as the task of learning linguistic and acoustic features of speech. Such models generate probability distribution of speech samples using pseudo-text units~\cite{hayashi2020discretalk,lakhotia2021generative, borsos2022audiolm}.  
%More recent works focus on modelling prosody~\cite{kharitonov2021text} and coarse-fine pseudo-text units~\cite{borsos2022audiolm}.
A speech language model generally consists of two main submodules: i) \emph{tokenizer} ($\text{Tok}$) and ii) autoregressive \emph{unit language model} (uLM) as shown in Fig.~\ref{fig:ulm}. 

A tokenizer takes continuous speech signal $\mathbf{x}$ as input and maps into a series of discrete units or tokens $\mathbf{d}$ as
\begin{equation}
    \text{Tok}(\mathbf{x}) = \mathbf{d} = \{d_1 \cdots d_T \},
\end{equation}
where each $d_i \in \{1 \cdots V\}$, $V$ is the size of vocabulary of such discrete tokens and $T$ is the total number of discrete tokens in $\mathbf{x}$, $T < t$. In general, it is similar to a tokenizer in a text language model. Tokenizer consists of an encoder (generally self-supervised-learning models) that extracts continuous features from speech $\mathbf{x}$: $\mathbf{f}=\{f_1,f_2 \cdots f_T\}$, where $f_i \in \mathbb{R}^{D}$ and $D$ is the dimension of encoded feature.   A quantizer then maps $\mathbf{f}$ into discrete speech unit tokens $\mathbf{d}$. 
For example GSLM~\cite{lakhotia2021generative} uses CPC~\cite{Riviere2021cpc}, HuBERT~\cite{hsu2021hubert} and wav2vec 2.0~\cite{baevski2020wav2vec} as feature encoder followed by k-means clustering~\cite{hartigan1979kmeans} as quantizer. Here the number of clusters ($k$) used in k-means clustering defines the unit vocabulary $V$. 
%For encoder, pretrained SSL and clustering models from other speech task as ASR are used. 
%V=k$. 

A speech unit language model is an autoregressive neural network that models the probability distribution over the set of discrete tokens $\mathbf{d}$ as
\begin{equation}
    p(\mathbf{d}) = \Pi_{i=1}^{T} p(d_i| d_{<i}).
\end{equation}
A uLM is trained to maximize the log-likelihood on a large speech dataset. Ideally, we can use any autoregressive network for the speech language model. Speech unit language models can generate human-like speech or continue a speech when fragment of speech is provided as context, similar to text generation capability of text-based language models as shown in~\cite{lakhotia2021generative}.
%A common choice is stacks of Transformer~\cite{vaswani2017transformer} decoder modules. 

%\subsubsection{Speech Discrete Units for GSLM}\label{ssec:sp_units}
\paragraph*{Speech discrete units for GSLM:}
In this work, we use Generative Spoken Language Modelling (GSLM) as the pretrained speech language model. 
%Choice of speech unit vocabulary can vary for speech language models. For example, 
GSLM uses small number of tokens: $V=50,100$ and $200$. Discrete units are generated from speech at a rate of 20~ms. Stacks of Transformer~\cite{vaswani2017transformer} decoder modules are used for language model network. Before unit language model training, GSLM removes repeated discrete tokens, i.e., the sequence $20,20,20, 16, 17, 17$ converts to $20, 16, 17$. The authors in ~\cite{lakhotia2021generative} hypothesize that such removal helps with limited attention span of Transformer network. Though such compression can help with reducing sequence length, it also looses useful information of speech token duration. Hence, we train a new language model to retain such information. We use the same Tokenizer as GSLM and train a new uLM without removal of repeated tokens. Since such repeated tokens increases the sequence length, we use a LSTM~\cite{hochreiter1997lstm} language model for less memory requirement.

%GSLM uses a small number of units, like $50$ or $100$ can model linguistic information but probably loses more fine acoustic details due to high compression of speech features $\mathbf{f}$ into a small number of tokens. More recent speech language models focus on modelling probability distribution over prosody and speech tokens~\cite{kharitonov2021text} or modelling coarse-fine discrete features using ~\cite{borsos2022audiolm}.
%More complex units can also be used like additional pitch information with small number of units~\cite{kharitonov2021text}) or large number of units~\cite{borsos2022audiolm}.  in the future we can extend the speech unit used in our work to more rich discrete tokens." at the end of this section

%%%%%%%%%%%%%%%%%%%%%%%%%%%%%%%%%%%%%%%%%%%%%%%%%%%%%%%%%%%%%%%%%%%%%%%%%%%%%%%%%%%%%%%%%%
\begin{table}[tb]
\caption{Utterance and system-level correlation with MOS in VoiceMOS challenge 2022 dataset (7106 files)~\cite{huang2022voicemos} with different configurations:  layer number ($L$) to extract feature from Hubert and number of clusters ($V$). We use SpeechLMScore with pretrained uLM. }
\label{tab:layers}
\centering
\begin{adjustbox}{width=\columnwidth,center}
\begin{tabular}{l | c c | c c c | c c c}
\hline
\toprule
&  & & \multicolumn{3}{c|}{\textbf{Utterance-level}} & \multicolumn{3}{c}{\textbf{System-level}} \\
ID & V & L & LCC & SRCC & KTAU & LCC & SRCC & KTAU\\
\midrule

50\_3 & \multirow{4}{*}{50} & 3 & \textbf{0.472} & 0.490 & 0.343 & 0.753 & 0.749 & 0.549\\
50\_4 & & 4 & 0.464 & 0.492 & 0.344 & \textbf{0.760} & \textbf{0.755} & \textbf{0.562} \\
50\_6 & & 6 & 0.462 & 0.462 & 0.321 & 0.694 & 0.692 & 0.496 \\
50\_12 & & 12 &  0.279 & 0.348 & 0.234 & 0.514 & 0.555 & 0.388 \\
\hline
100\_2 & \multirow{6}{*}{100} & 2 &  0.376 & 0.460 & 0.321 & 0.673 & 0.683 & 0.503 \\
100\_3 & & 3 &  0.322 & 0.505 & 0.355 & 0.598 & 0.666 & 0.490\\
%100 & 4 & 0.7 & 0.500 & 0.519 & 0.364 & 0.748 & 0.741 & 0.547 \\
100\_4 & & 4 &  0.379 & 0.527 & 0.370 & 0.705 & 0.741 & 0.552 \\
100\_5 & & 5 &  0.282 & 0.482 & 0.337 & 0.552 & 0.615 & 0.444 \\
%100 & 6 & 0.7 & 0.428 & 0.477 & 0.333 & 0.686 & 0.683 & 0.496 \\
100\_6 & & 6 & 0.300 & 0.454 & 0.317 & 0.523 & 0.559 & 0.392\\
%100 & 6 & 1.5 & 0.262 & 0.443 & 0.309 & 0.581 & 0.595 & 0.420\\
100\_12 & & 12 &  0.289 & 0.375 & 0.259 & 0.532 & 0.562 & 0.394 \\
\hline
200\_3 & & 3 & 0.419 & \textbf{0.538} & \textbf{0.380} & 0.719 & 0.726 & 0.539 \\
200\_4 &\multirow{2}{*}{200} & 4 & 0.464 & 0.536 & 0.378 & 0.701 & 0.700 & 0.511 \\
200\_6 & & 6 &  0.360 & 0.487 & 0.342 & 0.594 & 0.649 & 0.471\\
\bottomrule
\end{tabular}
\end{adjustbox}
\end{table}

\begin{table*}[tb]
\caption{Utterance and System-level correlation with MOS in VoiceMOS 2022 challenge testset (1066 files) and whole dataset (7106 files) with SpeechLMScore. }
\label{tab:bvcc}
\centering
\scalebox{0.8}{
\begin{tabular}{l | c c c | c c c | c c c | c c c}
\hline
\toprule
& \multicolumn{6}{c|}{\textbf{test-set}} & \multicolumn{6}{c}{\textbf{whole-set}} \\
& \multicolumn{3}{c|}{\textbf{Utterance-level}} & \multicolumn{3}{c|}{\textbf{System-level}} &  \multicolumn{3}{c|}{\textbf{Utterance-level}} & \multicolumn{3}{c}{\textbf{System-level}} \\
Model &LCC & SRCC & KTAU & LCC & SRCC & KTAU & LCC & SRCC & KTAU & LCC & SRCC & KTAU \\
\midrule
\multicolumn{13}{c}{Matched training domain} \\
\midrule
MOSNnet (pre) & 0.454 & 0.480 & 0.339 & 0.481 &  0.459 & 0.323 & 0.415 & 0.432 & 0.302 & 0.518 & 0.497 &  0.356\\
MOSNet (ft) &  \textbf{0.868} & \textbf{0.865} & \textbf{0.690} & \textbf{0.948} & \textbf{0.944} & \textbf{0.803}  & - &  - & - & - & - & -\\

\midrule
\multicolumn{13}{c}{Mismatched training domain} \\
\midrule
DNSMOS (SIG) &  \textbf{0.536} &  \textbf{0.553}  &  \textbf{0.392} &  \textbf{0.652} &  \textbf{0.684} &  \textbf{0.498} &  \textbf{0.495} &  \textbf{0.503} &  \textbf{0.354} &  \textbf{0.714} &  \textbf{0.720} &  \textbf{0.532} \\
DNSMOS (BAK) &  0.266 & 0.298 & 0.204 & 0.370 & 0.410 & 0.282 & 0.293 & 0.317 & 0.219 & 0.429 & 0.410 & 0.280 \\
DNSMOS (OVRL) & 0.496 & 0.497 & 0.352 & 0.606 & 0.623 & 0.450 & 0.473 & 0.473 & 0.334 & 0.678 & 0.668 & 0.488 \\
\midrule
\multicolumn{13}{c}{Unsupervised} \\
\midrule
SpeechLMScore (Pre) & 0.452 & 0.524 & 0.371 &  0.711 & 0.745 & 0.547 & 0.490 & 0.472 & 0.343 &  0.749 & \textbf{0.754} & 0.549 \\
%SpeechLMScore (Pre)(50\_4) ) &  0.403 & 0.518 & 0.366 & 0.726 & 0.742 & 0.542 & 0.464 & 0.492 & 0.344 & 0.760 & 0.755 & 0.562 \\
SpeechLMScore (LSTM) & 0.538 & 0.539 & 0.383 & 0.720 & 0.728 & 0.531 & 0.497 & 0.499 & 0.350 & 0.753 & 0.748 & 0.554 \\
%SpeechLMScore(LSTM)(50\_4) & 0.533 & 0.528 & 0.375 & 0.712 & 0.726 & 0.531 & 0.495 & 0.494 & 0.346 & 0.746 & 0.742 & 0.554 \\
SpeechLMScore (LSTM)+rep & 0.582 & 0.572 & 0.410 &  \textbf{0.743} &  \textbf{0.749} &  \textbf{0.551} &  \textbf{0.519} &  \textbf{0.516} &  \textbf{0.367} &  \textbf{0.759} &  0.739 &  \textbf{0.564} \\
%SpeechLMScore(LSTM)+rep(50\_4) & 0.586 & 0.410 & 0.381 & 0.731 & 0.744 & 0.545 & 0.533 & 0.524 & 0.370 & 0.748 & 0.734 & 0.547 \\
SpeechLMScore (LSTM) (Large) & 0.540 & 0.536 & 0.381 & 0.709 & 0.724 & 0.529 & 0.496 & 0.497 & 0.349 & 0.745 & 0.744 & 0.551 \\
%SpeechLMScore(LSTM)(50\_4) (Large) & 0.533 & 0.530 & 0.377 & 0.701 & 0.723 & 0.528 & 0.489 & 0.489 & 0.342 & 0.728 & 0.731 & 0.542 \\
SpeechLMScore (LSTM)+rep (Large) & \textbf{0.586} &  \textbf{0.584} & \textbf{0.419} & 0.729 & 0.736 & 0.539 & 0.514 &  \textbf{0.516} & 0.365 & 0.749 & 0.733 & 0.542 \\
\bottomrule
\end{tabular}
}
\end{table*}

%%%%%%%%%%%%%%%%%%%%%%%%%%%%%%%%%%%%%%%%%%%%%%%%%%%%%%%%%%%%%%%%%%%%%

\subsection{SpeechLMScore} 
\label{ssec:speech_lm_score}
As noted in Section~\ref{ssec:speech_lm}, uLM is trained to maximize the log-likelihood on a speech-only dataset. Thus a robust speech language model can be used to estimate the likelihood of a given speech. Here we propose that such a speech likelihood can also be used to evaluate the quality of speech. That is, given speech language model $\theta$, SpeechLMScore($\mathbf{x}| \theta$) is defined as
%We can then compute SpeechLMScore($\mathbf{x}| \theta$) as:
 \begin{equation}
 \text{SpeechLMScore}(\mathbf{d}|\theta) =\frac{1}{T} \sum_{i=1}^{T} \log p(d_i | d_{<i}, \theta).
 \end{equation}
More specifically, SpeechLMScore using uLM can be computed as follows: i) encode the speech into discrete tokens $\mathbf{d} = \{d_1 \cdots d_T\}$, and ii) for all tokens iteratively compute, log probability of current token $d_{i}$ provided all previous tokens $\{d_1 \cdots d_{i-1}\}$ using $\theta$, i.e., $\log p(d_i | d_{<i}, \theta)$ as shown in Fig.~\ref{fig:speech_lm_score}. 

SpeechLMScore measures the average log-probability of a set of speech tokens. Such metric is also related to perplexity of a speech sample. In fact, it is measuring how perplexed a speech language model is given set of discrete tokens from speech $\mathbf{x}$. Thus, lower perplexity, or higher log-likelihood, should correlate with human evaluations of higher speech quality. In the next section, we present experimental results demonstrating the effectiveness of the SpeechLMScore metric.

%%%%%%%%%%%%%%%%%%%%%%%%%%%%%%%%%%%%%%%%%%%%%%%%%%%%%%%%%%%%%%%%%%%%%%%%%%%%%%%%%%%%%%%%%%%%%%%%%%%%%%%%%%%%%%%%%%%%%%
\section{Experiments}
\label{sec:experiment}
In this section, we evaluate the generalizability of the proposed unsupervised metric through different speech evaluation tasks.

\subsection{Experimental setup}\label{ssec:exp_set}
%\subsubsection{Datasets}
\paragraph*{Dataset:}
Currently, the largest dataset to compare with human scores in speech synthesis/voice conversion is the VoiceMOS challenge~\cite{huang2022voicemos}. The main track of the challenge dataset contains a total of 7,106 utterances with 187 different systems and 38 samples per system. The dataset contains speech from three main sources: i) voice conversion challenges (VCC) from 2016, 2018 and 2020, a total of 79 systems~\cite{toda2016voice, lorenzo2018voice, yi2020voice}, ii) more recent text-to-speech models trained with  ESPnet~\cite{hayashi2020espnet} on LJspeech~\cite{panayotov2015librispeech}, 10 systems, and iii) older speech synthesis systems from past Blizzard challenges~\cite{king2008blizzard, kinga2009blizzard, prahallad2013blizzard} (2008-2016), a total of 98 systems. For each utterance, 8 human judgments are collected and the average score is used as ground truth human judgement. 
Next, we evaluate our approach on noisy speech and measure the correlation with the noisiness of speech using the Deep Noise Suppression challenge 2020~\cite{reddy2020dns} noisy reverberant testset. The testset contains 150 noisy samples, with varying signal-to-noise ratio (SNR) from 0-25 dB. Synthetic noisy speech contains clean speech from ~\cite{pirker2011pitch}, and 12 noise categories from AudioSet~\cite{gemmeke2017audioset}: fan, air conditioner, typing, door shutting, clatter noise, car, munching, creaking chair, breathing, copy machine, baby crying, and barking.

%\subsubsection{Speech unit language models}\label{ssec:ulm_setup}
\paragraph*{Speech unit language models:}
As discussed in section ~\ref{ssec:speech_lm}, We used pretrained models from GSLM~\cite{lakhotia2021generative} using fairseq\footnote{\url{https://github.com/facebookresearch/fairseq}}.  We use a pretrained tokenizer using \textsc{Hubert-base-ls960h} as the encoder and k-means clustering models as the quantizer. \textsc{Hubert-base-ls960h} consist of 12 Transformer encoder layers and trained with LibriSpeech 960 hour dataset~\cite{panayotov2015librispeech}. There are three versions of pretrained clustering models: $V$= $50,100$ and $200$ trained on LibriSpeech clean-100 hour subset. We use Hubert~\cite{hsu2021hubert} features and the corresponding clustering model as that was best performing uLM in the GSLM. Pretrained uLM consists of 12 Transformer decoder layers and trained using 3072 maximum sequence length and on a ``clean'' subset containing 6K hours speech~\cite{Riviere2021cpc} selected from LibriLight 60K dataset~\cite{Kahn2020librilight}. 
For our language model training, we use a LSTM language model. We train using two datasets: i) LibriLight medium segmented set with 5.6K hours of speech, and ii) another large dataset of approximately three times larger (16.8K hours) randomly selected from the LibriLight 60K hour dataset. We use a smaller subset from LibriLight 60K dataset for less computation requirement. Our LSTM language model contains 1024 hidden units with 3 layers and is trained with 0.2 dropout, 0.002 learning rate with Adam~\cite{kingma2014adam} optimizer for 40 epochs. For the larger dataset model (Large), we use a similar configuration with 4 layers. In this work, we report these speechLMscore models: SpeechLMScore (Pre): uses the GSLM pretrained model, SpeechLMScore (LSTM): uses our LSTM language model with removal of repeated tokens as GSLM, SpeechLMScore(LSTM)+rep: uses our trained LSTM model with repeated tokens, and SpeechLMScore(Large): language models trained on the larger subset.

%\subsubsection{Supervised Baselines}\label{ssec:baselines}
\paragraph*{Supervised baselines:}
We compare our unsupervised speechLMscore with two supervised MOS prediction models: MOSNet~\cite{reddy2021dnsmos} and DNSMOS~\cite{reddy2021dnsmos}. %Both of these models directly predict MOS in 1--5 scale. 
MOSNet is trained on the speech synthesis domain. The MOSNet pretrained model, MOSNet (Pre), is trained with the voice conversion 2018 challenge and MOSNet (ft) is finetuned on the VoiceMOS 2022 challenge dataset. DNSMOS is trained on the noise-suppression domain with ground truth human evaluation scores. DNSMOS  predicts three scores, signal quality (SIG), background quality (BAK), and overall quality (OVRL). We report the correlation with human evaluations and compare against supervised baselines using three metrics: Linear Correlation Coefficient (LCC)~\cite{pearson1920lcc}, Spearman Rank Correlation Coefficient (SRCC)~\cite{spearman1987srcc} and Kendall Tau (KTAU) rank correlation.

%%%%%%%%%%%%%%%%%%%%%%%%%%%%%%%%%%%%%%%%%%%%%%%%%%%%%%%%%%%%%%%%%%%%%%%%%%%%%%%%%%%%%%
\subsection{Experimental results}\label{sec:res}

\subsubsection{Analyzing best configuration on pretrained uLM}
Different layers of self-supervised features can be important for different tasks~\cite{hsu2021hubert, yang2021superb}. Another design choice for uLM is the number of tokens. Hence, we first compare different configurations of SpeechLMScore using pretrained GSLM. We mainly vary the layer number ($L$) of the Hubert model to extract features and the number of units ($V$). The results are reported in Table~\ref{tab:layers}. For this test, we use the whole dataset of the VoiceMOS 2022 challenge. We also varied temperature ($temp$) for computing probability but saw little to no effect and hence use $temp=1$ throughout. We observe lower layers were better correlated with human evaluations: for example $L=3$ and $4$ were consistently better than $L=12$. Also $50\_3, 50\_4, \text{~and~} 200\_3$ have highest correlation in different metrics. For the rest of the paper, we use $50\_3$ as the default SpeechLMScore tokenizer model, unless otherwise mentioned.
%%%%%%%%%%%%%%%%%%%%%%%%%%%%%%%%%%%%%%%%%%%%%%%%%%%%%%%%%%%%%%%%%%%%%
\begin{table}[tb]
\caption{Utterance-level and System-level correlation with MOS scores in different sources of VoiceMOS 2022 challenge dataset. VCC contains 3002 utterances, ES-TTS 380  and BLZ 3724.}
\label{tab:bvcc_domain}
\centering
\begin{adjustbox}{width=\columnwidth,center}
\begin{tabular}{ l | c | c c c | c c c }
\hline
\toprule
& & \multicolumn{3}{c|}{\textbf{Utterance-level}} & \multicolumn{3}{c}{\textbf{System-level}} \\
Source &  uLM & LCC & SRCC & KTAU & LCC & SRCC & KTAU\\
\midrule
VCC &  Pretrain & 0.505 & 0.501 & 0.355 & 0.863  & \textbf{0.829} & \textbf{0.643} \\
VCC &  LSTM+rep & \textbf{0.538} & \textbf{0.533} & \textbf{0.378} & \textbf{0.863}  & 0.820 & 0.625 \\
VCC &  LSTM (Large) & 0.484 & 0.482 & 0.338 & 0.842  & 0.811 & 0.616 \\
\hline
ES-TTS  & Pretrain & 0.404 & 0.288 & 0.201 & 0.632  & 0.44 & 0.333 \\
ES-TTS  & LSTM+rep & 0.296 & 0.243 & 0.169 & 0.576  & 0.358 & 0.289 \\
ES-TTS  & LSTM (Large) & \textbf{0.419} & \textbf{0.320} & \textbf{0.224} & \textbf{0.655}  & \textbf{0.527} & \textbf{0.422} \\
\hline
BLZ  & Pretrain & 0.268 & 0.321 & 0.219 & \textbf{0.487} & \textbf{0.557} & \textbf{0.384} \\
BLZ  & LSTM+rep & \textbf{0.323} & \textbf{0.328} & \textbf{0.226} & 0.481  & 0.506 & 0.360\\
BLZ  & LSTM (Large) & 0.316 & 0.314 & 0.215 & 0.480  & 0.529 & 0.367 \\
\bottomrule
\end{tabular}
\end{adjustbox}
\end{table}

\begin{table}[tb]
\caption{Utterance-level correlation with signal-to-noise ratio in DNS-challenge-2020 noisy reverb testset~\cite{reddy2020dns}}
\label{tab:noisy_dns1}
\centering
\scalebox{0.85}{
\centering
\begin{tabular}{l |  c c c  }
\hline
\toprule
Model &  LCC & SRCC & KTAU \\
\midrule
\multicolumn{4}{c}{Matched training domain} \\
\midrule
DNSMOS & 0.349 & 0.352 & 0.248 \\ 
\midrule
\multicolumn{4}{c}{Mismatched training domain} \\
\midrule
MOSNet (Pre) & 0.286 & 0.324  & 0.221 \\
MOSNet (ft) & 0.443 & 0.456 & 0.325 \\
\midrule
\multicolumn{4}{c}{Unsupervised} \\
\midrule
%SpeechLMScore (Pre) & 0.448 & 0.437 & 0.310 \\
SpeechLMScore (Pre) & 0.448 & 0.437 & 0.310 \\
SpeechLMScore (Pre) (50\_4) & 0.475 & 0.496 & 0.339 \\
SpeechLMScore (LSTM) & 0.477 & 0.461 & 0.332 \\
SpeechLMScore (LSTM) (50\_4) & \textbf{0.538} & \textbf{0.518} & \textbf{0.374} \\
SpeechLMScore (LSTM) + rep & 0.452 & 0.433 & 0.308 \\
SpeechLMScore (LSTM) + rep (50\_4) & 0.478 & 0.455 & 0.324 \\
SpeechLMScore (LSTM) (Large) & 0.485 & 0.472 & 0.340 \\
SpeechLMScore (LSTM) (Large) (50\_4) & 0.527 & 0.506 & 0.364 \\
\bottomrule
\end{tabular}
}
\end{table}

%%%%%%%%%%%%%%%%%%%%%%%%%%%%%%%%%%%%%%%%%%%%%%%%%%%%%%%%%%%%
\subsubsection{Correlation with Voice Conversion and TTS}\label{ssec:voicemos}
We report correlation with human evaluation on the VoiceMOS 2022 challenge testset of $1066$ files and for all files $7106$ files in Table~\ref{tab:bvcc}. We compare with two baselines: MOSNet, which is trained on the task and dataset and DNSMOS, trained on a separate task. Since MOSNet (ft) is trained using 4974 utterances (training set) from the VoiceMOS dataset, we only include MOSNet (ft) results on the testset. MOSNet finetuned on the challenge dataset has the highest correlation to the testset. With DNSMOS scores, the signal quality score has the highest correlation with naturalness MOS. SpeechLMScores show higher correlation than all DNSMOS scores but lower than MOSNet. Furthermore, SpeechLMScore with duration information of tokens improve correlation.  SpeechLMScore shows encouraging correlations regardless of not using target scores at all.  %shows unsupervised scores have room for further improvement compared to supervised models.

Next, we analyze the correlation on the three distinct sources of the VoiceMOS dataset: voice conversion challenges (VCC), ESPnet-tts (ES-TTS), and Blizzard challenges (BLZ) as shown in Table~\ref{tab:bvcc_domain}. We observe SpeechLMScore is highly correlated with human evaluation in voice challenge datasets, followed by both TTS systems. However, among the TTS systems, we observe, from the Blizzard challenge samples, the lowest correlation with older TTS systems. Upon further inspection, we observe SpeechLMScore to have bias towards speech fluency more than pitch characteristics which is the main factor of evaluation in older TTS models~\cite{lemaguer22_interspeech}. 

% Results are reported in Table~\ref{tab:bvcc_all}. Compared to pretrain model we observe improvement in correlation in both LSTM language model, we observe best performance using LSTM language model and with repeated tokens. 

\subsubsection{Correlation with noise in speech}\label{ssec:dns}
Next, we evaluate if SpeechLMScore correlates with the presence of noise in speech. For this experiment, we use Deep Noise Suppression (DNS) challenege-2020~\cite{reddy2020dns} noisy synthetic testset with reverberation. We measure the correlation with signal-to-noise ratio of the speech samples and report in Table~\ref{tab:noisy_dns1} with both supervised baseline MOSNet and DNSMOS. We observe SpeechLMScore achieves a higher correlation with signal-to-noise ratio than both DNSMOS and MOSNet. Therefore, we conclude that such automatic unsupervised metric can generalize to different speech evaluation tasks.

\section{Conclusions}\label{sec:conc}
We propose SpeechLMScore, an automatic metric for evaluating speech samples using speech language models. Proposed metric is simple, easy to use and does not require reference speech sample. Furthermore, SpeechLMScore is trained using speech dataset only, and does not need large-scale human evaluation data. We show our proposed metric is highly correlated with human evaluation in different speech systems like voice conversion, text-to-speech using the VoiceMOS challenge dataset. We also show SpeechLMScore correlates  with SNR using DNS 2020 challenge testset.
Though proposed metric is generalizable to different tasks, there is scope for further improvement. Future extensions are envisioned to use more complex and prosody-rich tokens. 

\paragraph*{Acknowledgements:}
This work used the Extreme Science and Engineering Discovery Environment (XSEDE) ~\cite{xsede}, which is supported by National Science Foundation grant number ACI-1548562. Specifically, it used the Bridges system ~\cite{nystrom2015bridges}, which is supported by NSF award number ACI-1445606, at the Pittsburgh Supercomputing Center (PSC).

\vfill\pagebreak
\newpage
\footnotesize
\bibliographystyle{IEEEbib}
\bibliography{refs}

\end{document}